\documentstyle[preprint,aps,floats,epsfig]{revtex}
\tighten
\begin{document}
\draft
\title{Intrinsic Broadening of the Transverse Momentum Spectra in
  Ultrarelativistic Heavy-Ion Collisions?}
\author{Larry McLerran$^a$ and J\"urgen Schaffner-Bielich$^b$} 

\address{$^a$Department of Physics, Brookhaven National Laboratory, 
Upton, NY 11973-5000, USA \\
$^b$RIKEN BNL Research Center, Brookhaven National Laboratory, 
Upton, NY 11973-5000, USA}
\date{\today}

\maketitle   

%%%%%%%%%%%%%%%%%%%%%%%%%%%%%%%%%%%%%%%%%%%%%%%%

\begin{abstract} The mean transverse momentum in high energy nuclear
collisions is studied. The formation of a Color Glass Condensate results
in an increase of the square of the mean transverse momentum with the
charged multiplicity per unit rapidity and unit transverse area. This
prediction is in agreement with the presently available data, in
particular with the heavy-ion data, lending support to the notion that the
transverse momentum spectra in ultrarelativistic collisions of heavy
nuclei might be  controlled by intrinsic $p_T$ broadening and not by
hadronic rescattering, i.e.\ transverse flow. \end{abstract}

\pacs{}

%%%%%%%%%%%%%%%%%%%%%%%%%%%%%%%%%%%%%%%%%%%%%%%%

The initial gluon distribution in high-energy nuclear collisions can be
described by the formation of a so called Color Glass Condensate (CGC) 
\cite{mv,jkmw,iancu00}.  The gluon multiplicity per unit rapidity and
transverse area has been shown to depend on the initial gluon density
which saturates at very high energies \cite{Gribov,MuellerQiu,kmw,alex00,kn}. 
This saturation should be observable in the mean transverse momentum spectra
which should scale then like the square root of the charged multiplicity
per unit rapidity and unit {\em transverse} area, hence:  
\begin{equation}
   \langle p_T \rangle^2 \; \sim \;  \frac{1}{\pi R^2} \frac{dN_c}{dy}
\label{eq:scaling}
\end{equation} 
by dimensional arguments.  In this paper, we are going to
confront this prediction with presently available experimental data.
As we will outline in the following, we find that the above relation
eq.~(\ref{eq:scaling}) is fulfilled by both, pp or p$\bar{\rm p}$ and heavy-ion
data. This observation seems to suggest that the transverse momentum
broadening seen in elementary and heavy-ion collisions results from the
same underlying physics, namely the intrinsically generated $p_T$
broadening in the partonic phase, i.e.\ from the CGC. This implies that
transverse flow from rescattering in heavy-ion collisions in the hadronic
phase might be small compared to the transverse push initially generated
in the partonic phase. 

\begin{figure}[t]
\centerline{\epsfig{figure=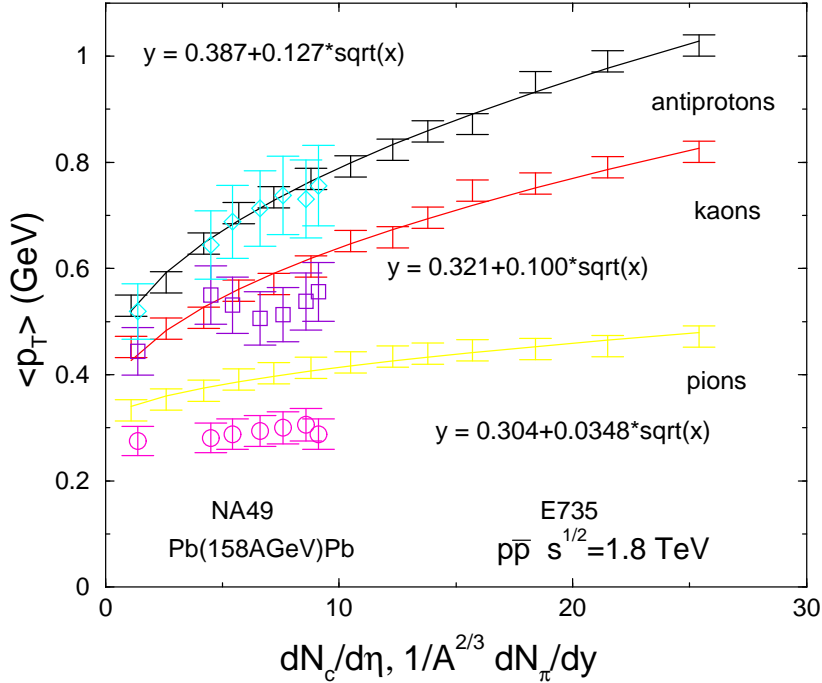,height=0.4\textheight}}
\caption{The mean $p_T$ as a function of the charged multiplicity per unit 
  rapidity and area. Shown are the data points for p$\bar{\rm p}$ collisions at
  $\sqrt{s}=1.8$ TeV from E735 at the Tevatron as taken from \protect\cite{e735_93},
  and heavy-ion data for pp and PbPb collisions at 158 AGeV from NA49 at the SPS as
  extracted from \protect\cite{na49_qm99}. The solid lines are fits through the
  Tevatron data using the scaling law of the CGC.}
\label{fig:ptmult}
\end{figure}

The study of the particle transverse momentum as a function of multiplicity is interesting
as it might reveal features of a phase transition to a quark-gluon plasma 
\cite{vanHove}. 
The increase of the mean $p_T$ with the charged multiplicity is a well
known experimental fact for high-energy electron and proton-antiproton collisions (see
e.g.\ \cite{ua1,e735_88} and references therein). 
Mass identified particle production were measured at the Tevatron
for p$\bar{\rm p}$ collisions at 
up to $\sqrt{s}=1.8$ TeV by the E735 collaboration \cite{e735_90,e735_93}. 
The data is shown in fig.\ \ref{fig:ptmult}. 
It is apparent from the figure, that the increase of the mean $p_T$ 
depends on the mass of the hadron and is stronger for heavier particles.
We have also plotted the results of 
two-parameter fits using the scaling law
eq.~(\ref{eq:scaling}) by the solid curves.
The two parameters are the coefficient of $\sqrt{dN_c/dy}$ and
the value of $p_T$ as the multiplicity shrinks to zero. 
We note that the agreement with
the data is quite reasonable. 
The mean $p_T$ seems to follow the saturation
momentum of the gluon distribution function resulting in a square root
dependence with the multiplicity density. 
The onset of the curve at small mean $p_T$ is due to soft
physics. The difference in the slope parameter for pions, kaons, antiprotons
will be discussed later. 

We have also checked numerically that the more accurate data for all
negatively charged particles as measured by the UA1 collaboration is in
accord with the scaling law (\ref{eq:scaling}).  The slope is slightly
larger than for the pions in fig.\ \ref{fig:ptmult} due to the small
admixture of negatively charged kaons and antiprotons.
The increase of the mean $p_T$ with multiplicity for p$\bar{\rm p}$
for the different hadrons has been also reproduced by more sophisticated
partonic models, like 
the HIJING model \cite{wang}, 
the quark-gluon string model \cite{qgsm},
and the dual parton model \cite{dpm}.
Radial flow has been also suggested as a possible explanation \cite{levai},
but it is at variance with the small source sizes determined from two-pion
correlations \cite{e735_corr}.

Next, we test our model for relativistic (heavy-)ion
collisions. Presently published data with the highest
available bombarding energy is from the ISR collider at CERN for light
ions up to $\alpha\alpha$ collisions \cite{isr} 
An enhancement of the
slope of the mean $p_T$ with the multiplicity density is observed. At
the same bombarding energy of $\sqrt{s}=31$ GeV, the increase amounts to
about 30\% for $\alpha\alpha$ collisions compared to dd collisions, which
is about 30\% larger than that for pp collisions. According to
(\ref{eq:scaling}), the increase of the slope of the mean $p_T$ squared
should scale like the unit transverse area, i.e.\ with $A^{2/3}$. More
precisely, it would scale like $r_p^2/(r_0^2 A ^{2/3})$ where $r_p$ is the proton
radius and $r_0\approx 1.1$.
The mean $p_T$ will increase then by about $2^{1/3}$, i.e.\ by 26\%, when
going from pp to dd to $\alpha\alpha$ in excellent agreement with the
observation made at ISR \cite{isr}. 

Concerning truly heavy-ion collisions, the first events from Brookhaven's
Relativistic Heavy-Ion Collider (RHIC) for AuAu collisions at
$\sqrt{s}=130$ AGeV are presently under investigation \cite{barbara}.  The
NA49 experiment at CERN's SPS has published multiplicity and
mean $p_T$ distributions as a function of the centrality for PbPb
collisions at midrapidity \cite{na49_qm99}. We extract the charged particle
multiplicity per unit rapidity from the pion multiplicity and by setting
$dN/dy=160$ for the most central bin. The charged multiplicity density per
unit transverse area is then given by dividing it with $A^{2/3}$ (setting the
proton radius to $r_p=1$ fm would give an additional shift in the scale down
by $(r_p/r_0)^2$, i.e.\ by 20\%). 
Systematic error bars for the mean $p_T$ have not been given in
\cite{na49_qm99}. We adopt an ad hoc value of 10\% for the systematic
error \cite{sikler}. The resulting points are plotted in fig.\
\ref{fig:ptmult}. We note that the results for the protons and kaons are
then within the error bars on top of the data from the Tevatron and the
corresponding fit for a CGC. The results for the mean $p_T$ for the
pions are somewhat lower though. 

However, this difference is due to the moderate dependence of the mean
$p_T$ on the bombarding energy (see \cite{ua1,e735_93} and references
therein). For energies comparable to the SPS the mean transverse momentum averaged over
multiplicity $\langle p_T \rangle$ is about 0.33 GeV for pions for p$\bar{\rm p}$ collisions and
increases to 0.38 GeV at $\sqrt{s}=1.8$ TeV. If the points for the pions
from NA49 are shifted accordingly, they are consistent with our fit to the
Tevatron data points. A similar shift for the kaons points does not change
the agreement with the scaling curve. The proton points are then slightly
higher than our scaling curve fitted to the Tevatron data, but they
are consistent with the Tevatron data points within the error bars.  The slight energy
dependence of $\langle p_T \rangle$ can be explained by a slight increase of the
saturation momentum with $\sqrt{s}$. Assuming a power law behavior
\cite{Yuri}, we find that the energy dependence of $\langle p_T \rangle$ for charged
particles as measured by the UA1 collaboration goes like $\sim E^\delta$
\cite{ua1} with $\delta=0.08$. Note, that a logarithmic dependence is also
in accord with the data \cite{ua1}. The increase with energy is
considerably larger for heavier hadrons as seen by the E735 collaboration
\cite{e735_93}, so that for protons we get $\delta=0.19$ instead. We
remark, that the data seems to indicate that there is an threshold effect,
an onset of the increase of the transverse momentum at an energy around
$\sqrt{s}=100-200$ GeV which is right in the region accessible at RHIC. 
The gluon distribution function starts to turn on at $x<0.01$.
This is the region where the rise in $p_T$ should begin. In heavy-ion reactions
the start of the rise appears at smaller energies which might be due to the
increase of the gluon density in AA collisions compared to pp collisions. 
The onset of gluon saturation might be visible by studying the mean transverse 
momentum for pA and light AB systems at the SPS. At a certain size of the
colliding system, the $\langle p_T \rangle^2$ dependence should then increase with the 
multiplicity per unit transverse area. There will be also  
a moderate enhancement of $\langle p_T \rangle$ compared to pp collisions due to 
the increase of the saturation momentum $Q_s^2$ with $A^{1/3}$.
The A dependence of the color-glass condensate for pA and light systems has
been studied in detail in \cite{Adrian2001}. 

In conclusion, there seems to be a dramatic correspondence of the mean
$p_T$ appropriately 
scaled by the transverse unit area for heavy-ion collisions and p$\bar{\rm p}$
collisions at much larger energies. The origin of this behavior is lying in
the saturation of the initial gluon transverse momentum distribution. This
observation, if taken seriously, leads to a drastic conclusion: it is possible 
that the transverse momentum
distribution of hadrons in heavy-ion collisions is mainly controlled by the
initial momentum distributions of gluons and not by effects from hadronic
rescattering. The good agreement between the two curves from p$\bar{\rm p}$ and
heavy-ion collisions for the mean $p_T$ does not leave room for a
significant transverse expansion via a collective transverse flow.
We note, that a random walk model can also describe the $p_T$ broadening in
heavy-ion collisions without transverse flow \cite{Leonidov} 
(see also \cite{Satz2000} for a critical discussion).

What we left open is the question why heavier particles have larger mean
$p_T$ in p$\bar{\rm p}$ and in heavy-ion collisions. A possible explanation can be
provided by the coalescence model for the formation of hadrons at the
confinement phase transition. 
Of course, the formation of hadrons is an
extremely complex process which requires a detailed dynamical approach. 
With this in mind, let us state the following.
Naively, the mean $p_T$ of a hadron will scale
with the number of quarks in the coalescence model as the quark momenta are
added up to form a hadron. 
The mean $p_T$ for protons should then be 3/2 times
larger than that for pions. 
The strange quark will take a special role in this 
picture as the associated production from gluons needs to overcome the mass threshold.
After coalescence, many pions will come from subsequent resonance decays, e.g.\ from
the $\rho$ meson. The $\rho$ meson picks up 2 times the intrinsic $p_T$, but
decays to two pions with about half that $p_T$. Hence, the combination of
these effects results in a $p_T$ ratio between 3/2 and 3. The momentum spectra 
of kaons on the other hand are not affected by this resonance effect and end up 
with about 2/3 times the proton mean $p_T$. 
Indeed, the mean $p_T$ for protons compared to pions shows
an increase by about 1.5--2.0 over a wide range of energy (see e.g.\ the data compiled in
\cite{e735_93}), the kaon mean $p_T$ being between the mean $p_T$ for pions and 
protons. Also, the slope parameters measured for pions and protons in
PbPb collisions has been extracted to be around 190 MeV and 300 MeV
\cite{na49_slope}, respectively, reflecting a similar increase in the mean $p_T$.

We conclude with a brief comment about a possible test of this picture.
The physical quantity which is computed from the Color Glass Condensate is 
the phase space distribution 
\begin{equation}
        {1 \over {\pi R^2}} {{dN} \over {dyd^2p_T}} \quad .
\end{equation}
On dimensional grounds for equal size nuclear collision at some fixed
impact parameter
\begin{equation}
        {1 \over \sigma} {{dN} \over {dyd^2p_T}} = F(Q_s^2/p_T^2) \quad .
\end{equation}
In this equation, $Q_s^2$ is the saturation momentum squared which
characterizes the Color Glass Condensate at the impact parameter for the
collision.  An attempt to extract this saturation momentum has been done
by Kharzeev and Nardi \cite{kn}.  The cross section $\sigma$ is that
for a collision to occur at the impact parameter at hand. 
This formula predicts that the $p_T$ distributions at different impact
parameters can be rescaled into one another.  If true, it is a remarkable
verification of the approximate scale-invariance of the Color Glass
Condensate.

%%%%%%%%%%%%%%%%%%%%%%%%%%%%%%%%%%%%%%%%%%%%%%%%%
 
\acknowledgments

J.S.B. thanks the Institut f\"ur Theoretische Physik at the University of Frankfurt,
where parts of this work were completed, for their warm hospitality,
and RIKEN, BNL, and the U.S. Department of Energy for providing the
facilities essential for the completion of this work. 
This manuscript has been authorized with the U.S. Department of Energy under
Contract No. DE-AC02-98CH10886. 

%%%%%%%%%%%%%%%%%%%%%%%%%%%%%%%%%%%%%%%%%%%%%%%%%


\begin{thebibliography}{99}

\bibitem{mv} 
L. McLerran and R. Venugopalan, Phys. Rev. {\bf D49}, 2233 (1994); {\it ibid} 3352 (1994); 
Y.~Kovchegov, Phys. Rev.  {\bf D54}, 5463 (1996).

\bibitem{jkmw} 
J. Jalilian-Marian, A. Kovner, L. McLerran, and H. Weigert, 
Phys. Rev. {\bf D55}, 5414 (1997); 
J. Jalilian-Marian, A. Kovner, A. Leonidov, and H. Weigert, 
Nucl. Phys. {\bf B504}, 415 (1997); Phys. Rev. {\bf D59}, 014014 (1999).

\bibitem{iancu00}
E. Iancu, A. Leonidov, and L. McLerran, hep-ph/0011241  (2000).

\bibitem{Gribov}
L.~V. Gribov, E.~M. Levin, and M.~G. Ryskin, Phys. Rep. {\bf 100},  1  (1983).

\bibitem{MuellerQiu}
A.~H. Mueller and J.~W. Qiu, Nucl. Phys. {\bf B268},  427  (1986).

\bibitem{kmw} 
A. Kovner, L. McLerran, and H. Weigert, 
Phys. Rev. {\bf D52}, 6231 (1995); {\it ibid} 3809 (1995). 

\bibitem{alex00}
A.~Krasnitz and R.~Venugopalan, Phys. Rev. Lett.  {\bf 86}, 1717 (2001);
W. P\"oschl and B. M\"uller, Comput. Phys. Commun. {\bf 125}, 282 (2000).

\bibitem{kn}  
D. Kharzeev and M. Nardi,  Phys. Lett. B {\bf 507}, 121 (2001).

\bibitem{vanHove}
L. van Hove, Phys. Lett. {\bf 118B}, 138 (1982).

\bibitem{ua1}
UA1 collaboration, C. Albajar {\it et~al.}, Nucl. Phys. {\bf B335},  261  (1990).

\bibitem{e735_88}
E735 collaboration, T. Alexopoulos {\it et~al.}, Phys. Rev. Lett. {\bf 60},  1622  (1988).

\bibitem{e735_90}
E735 collaboration, T. Alexopoulos {\it et~al.}, Phys. Rev. Lett. {\bf 64},  991  (1990).

\bibitem{e735_93}
E735 collaboration, T. Alexopoulos {\it et~al.}, Phys. Rev. {\bf D48},  984  (1993).

\bibitem{wang}
X.-N. Wang and M. Gyulassy, Phys. Rev. {\bf D45},  844  (1992).

\bibitem{qgsm}
N.~S. Amelin, E.~F. Staubo, and L.~P. Csernai, Phys. Rev. {\bf D46},  4873
  (1992).

\bibitem{dpm}
F.~W. Bopp, R. Engel, D. Pertermann, and J. Ranft, Phys. Rev. {\bf D49},  3236
  (1994).

\bibitem{levai}
P. Levai and B. M{\"u}ller, Phys. Rev. Lett. {\bf 67},  1519  (1991).

\bibitem{e735_corr}
E735 collaboration, T. Alexopoulos {\it et~al.}, Phys. Rev. {\bf D48},  1931  (1993).

\bibitem{isr}
A. Breakstone {\it et~al.}, Phys. Lett. {\bf B183},  227  (1987).

\bibitem{na49_qm99}
NA49 collaboration, J. B\"achler {\it et~al.}, Nucl. Phys. {\bf A661},  45  (1999).
 
\bibitem{barbara}
J.~Velkovska [PHENIX collaboration], nucl-ex/0105012 (2001);
J. W. Harris and M. Calderon [STAR collaboration], 
talks given at Quark Matter 2001, Stony Brook, New York, 2001; 
Barbara Jacak, private communication. 

\bibitem{sikler}
Ferenc Sikl\'er, private communication.

\bibitem{Yuri}
Y.~V. Kovchegov, Phys. Rev. {\bf D61},  074018  (2000).

\bibitem{Adrian2001}
A.~Dumitru and L.~McLerran, hep-ph/0105268 (2001).

\bibitem{Leonidov}
A. Leonidov, M. Nardi, and H. Satz, Z. Phys. {\bf C74},  535  (1997).

\bibitem{Satz2000}
H.~Satz, Nucl. Phys. B (Proc. Suppl.) {\bf 94}, 204 (2001).

\bibitem{na49_slope}
NA49 collaboration, T. Wienold {\it et~al.}, Nucl. Phys. {\bf A610},  76c  (1996).

\end{thebibliography}
\end{document}